\documentclass[twocolumn,a4paper]{article}

\def\bbox#1{{\bf #1}}
\def\text#1{\mbox{#1}}

\usepackage{epsfig}

\begin{document}

\twocolumn[%
\begin{center}
\vspace*{-3ex}
\mbox{
Extended abst. subm. to   
{\em Electron Transport in Mesoscopic Systems}
(G\"oteborg, Sweden, satellite to LT22)
}

\textbf{\Large
Quasiclassical theory of superconductivity: \\
a  multiple interface geometry
}\vskip1ex
{A. Shelankov$^{1,2}$ and M. Ozana$^{1}$}
\vskip0.5ex
\textsl{
$^{1}$Department of Theoretical Physics, Ume{\aa} University, 901 87,
Ume{\aa}, Sweden\\
$^{2}$A. F.  Ioffe Physico-Technical
Institute, 194021 St.Petersburg, Russia
}
\end{center}
]

In many cases of interest such as a multi-layer mesoscopic structure
(e.g. a superlattice) or the grain boundaries network in high-T$_c$'s,
one deals with the situation where electrons traverses many partially
transparent interfaces without loosing coherence.  Then, the
consequent reflection/transmission events require a simultaneous
consideration.  Theoretically, even an isolated interface poses
certain difficulties: Since abrupt changes violate the quasiclassical
condition, the theory of superconductivity in terms of the
quasiclassical matrix Green's function $\hat{g}^{R}$ \cite{quasi} is
invalid at interfaces.  The interface is included via the boundary
condition derived by Zaitsev \cite{Zai84} -- a cubic matrix relation.
In case of many interfaces, one comes to a system of nonlinear matrix
equations solution of which is far not simple if possible.  Moreover,
some authors argue \cite{Nagai} that the normalization condition is
violated in the multiple interface case so that the quasiclassical
scheme fails.  The purpose of this paper is to suggest a new method
which allows one to study systems with multiple partially transparent
interfaces $SIS'$ or $SIN$ in the framework of the quasiclassical
theory of superconductivity.  Motivated by the paramagnetic effect
problem (see references in \cite{FauBelBla99}), the response to the
vector potential (the superfluid density) of a $ISIS'I$ sandwich is
calculated.

A convenient starting point is the formulation \cite{She85} of the
quasiclassical technique in terms of the 2-point trajectory Green's
function $\hat{g}^{R}(x_{1},x_{2})$ where $x_{1,2}$ are the
coordinates along the trajectory characterised its direction
$\bbox{n}$ and position $\bbox{R}\perp \bbox{n}$.  The equation of
motion reads \vspace{-2.5ex}
\begin{equation}
(i v {\partial\over{\partial x_{1}}} + \hat{H}^{R}(x_{1}))
\hat{g}^{R}(x_{1}, x_{2})
      =   i v \delta (x_{1}-x_{2})
\label{78a}
\end{equation}

\vspace{-2.5ex} 
\noindent 
where $ \hat{H}^{R} =
 \left({ \varepsilon \atop \Delta^{*}}
 {\Delta\atop - \varepsilon }\right)- \Sigma^{R}$ and the self-energy
is denoted $\Sigma^{R}$. 
The Green's function $\hat{g}_{\bbox{p_{F}n}}^{R}$ of Ref. \cite{quasi}
is found by $\hat{g}^{R}(x\pm 0,x) = (\hat{g}_{\bbox{p_{F}n}}^{R}\pm
1)/2$.

The key point is the understanding that in the presence of partially
reflective interfaces a {\em typical} trajectory remains extended and
simply connected.  In other words, the trajectories typically do not
have loops and are not split to interfering paths so that there is
only one path connecting any two points on the trajectory.  Then, the
coordinate along the trajectory is well defined and the method of
2-point Green's function is applicable.  Due to
reflection/transmission at the interfaces, the typical path becomes
tree-like with ``knots'', {\it i.e.}  the points where ballistic
pieces of the trajectory cross.  In between knots, the quasiclassical
equation Eq.(\ref{78a}) is valid, whereas the knots are included via
the boundary matching condition.

The latter is formulated for the ``wave functions'' $\phi =
{u\choose v}$, 
$\bar{\phi }\equiv (v,-u)$ 
factorizing
$\hat{g}^{R}$, \rule[-1.5ex]{0ex}{0ex} 

\centerline{$
\rule[-1.5ex]{0ex}{0ex} \hat{g}^{R}(x +0, x) = \phi_{+}(x) \bar{\phi
}_{-}(x)$,}

found from the linear equations, 

\vspace{-4ex}

\begin{equation}
(i v {\partial\over{\partial x}} +\hat{H}^{R}(x))\phi= 0
\;, \phi_{\pm}(x \rightarrow \pm \infty)=0,
\label{08a}
\end{equation}

\vspace{-2ex}

\noindent 
and normalized, $\bar{\phi}_{-}\phi_{+}=1$.

A general interface is partially diffusive, mixing states within a
continuum of momenta. As a model, one allows mixing a finite number of
states, $N $ incoming and $N$ outgoing, at a knot.  The knot value of
the wave functions $\phi_{i}$ ($\phi_{k'}$) in the in-coming
(out-going) channels, $i,k' = 1,\ldots,N$ are related to each other by
the unitary scattering matrix \cite{SheOza}

\vspace{-1.8ex}
\begin{equation}
\phi_{k'} = \sum\limits_{i=1}^{N} S_{k'i}\phi_{i} \; .
\label{98a}
\end{equation}

\vspace{-1.5ex}

\noindent 
The energy independent matrix $S$, the normal state property, is
considered as a given input.

In this scheme, one finds the ``wave function'' $\phi $ by solving
Eq.(\ref{08a}) on the pieces of the trajectory in between the knots
and uses Eq.(\ref{98a}) to tailor the pieces.  The scheme allows one
to consider the usual and Andreev reflections on equal footing,
simultaneously with (in)elastic scattering processes included via the
self-energy $\Sigma^{R}$.

\mbox{Here we restrict ourselves to the simplest case of} 
\parbox{80pt}{
 \epsfig{file= 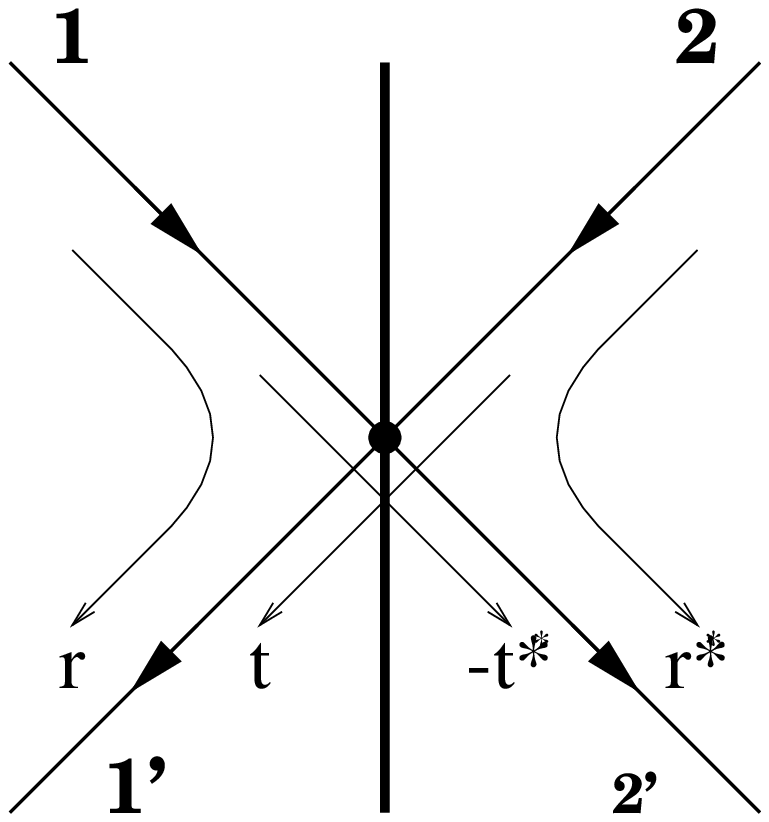,height=75pt,width=90pt,angle=0}
}
\hfill\hfill 
\begin{minipage}{4cm}\
$N=2$
(corresponding
e.g. to a partially transparent specular interface) when the S-matrix
takes the form $\left({r\atop -t^{*}}{t\atop r^{*}} \right)$,
$R=|r|^{2}$, $T= |t|^{2}$, $R+T=1$.  \end{minipage}

Analysis \cite{SheOza} shows that Eq.(\ref{98a}) leads to the
following relation among the knot values of the 1-point Green's
functions $\hat{g}^{R}_{1}$ and $\hat{g}^{R}_{1'}$ on the trajectories
1 and $1'$:
\begin{equation}
\hat{g}_{1'}^{R} =
{\cal M}
\hat{g}_{1}^{R}
{\cal M}^{-1}
\label{a9a}
\end{equation} 
where the transfer matrix ${\cal M}$ reads
\begin{equation}
{\cal M}= {1\over 2r^{*}}(1+R)
\left(1 - {T\over 1+R} \hat{g}_{2' \bullet 2}^{R}\right)
\; ,
\label{b9a}
\end{equation}
here $\hat{g}_{2' \bullet 2}^{R}$, the ``across the interface'' propagator, is
\begin{equation}
\hat{g}_{2'\bullet 2}^{R}= {1\over 1 +
{1\over 2}[\hat{g}_{2'}^{R},\hat{g}_{2}^{R}]_{+}}
\left(\hat{g}_{2'}^{R}+ \hat{g}_{2}^{R}+ {1\over
2}[\hat{g}_{2'}^{R},\hat{g}_{2}^{R}]_{-}
\right)\;\; ,
\label{c9a}
\end{equation}
where $[...]_{(+)-}$ denotes (anti)commutator.

Eqs.(\ref{a9a}-\ref{c9a}) being equivalent to Zaitsev's boundary
conditions, seem to be more convenient for numerical
calculations. Besides, they can be easily generalised \cite{SheOza} to
the case of a diffusive interface modelled by a knot $N>2$.

To illustrate the method, we calculate the density of states and the
superfluid density for a system with 3 planes of reflection: a
sandwich made of two superconducting layers $L$ and $R$ separated by a
partially transparent interface; the layers occupy the regions $-d_{L}
<z<0$ and $0< z<d_{R}$ and the order parameter is $\Delta_{L}$ and
$\Delta_{R}$, respectively.

Consider the periodic zig-zag trajectory in the R-layer, formed by
reflections from the boundary at $z=d_{R}$ and the interface at $z=0$;
it is specified by the angle $\theta $ and the period $a_{\theta
}=2d_{R}/\cos\theta $.  Knots are the points where the zig-zag touches
the interface.  Because of the periodicity, each of the regular
solutions $\phi_{\pm}(x)$ is proportional to one of the eigenvectors
of the evolution matrix $\hat{U}_{x}$ defined via $\phi (x + a_{\theta
})= \hat{U}_{x}\phi (x)$. The evolution matrix reads,

\vspace{-1.5ex}
\begin{equation}
\hat{U}_{x}= \hat{U}^{(0)}(x,0){\cal M}\hat{U}^{(0)}(0, x-a_{\theta })
\; ,
\label{d9a}
\end{equation}

\vspace{-1.2ex}
\noindent 
where ${\cal M}$ is the transfer matrix Eq.(\ref{b9a}), and
 $\hat{U}^{(0)}$ is the bulk evolution operator corresponding to
Eq.(\ref{08a}): $\hat{U}^{(0)}(x+y,y)= \exp [i x \hat{H}^{R}/v]$.
One can show \cite{SheOza} that the 1-point Green's function for the
 direction $n_{z}= \pm \cos\theta $ can be
calculated as

\vspace{-2ex}
\begin{equation}
\hat{g}^{R}(x) =
\hat{U}_{x}'\left/\sqrt{\left(\hat{U}_{x}'\right)^{2}}\right.
\; ,
\label{e9a}
\end{equation}

\vspace{-3ex}
\noindent 
where 
$\hat{U}_{x}'$ stands for $\hat{U}_{x} - \left({1\over 2}{\rm
Tr}\,\hat{U}_{x}\right) \hat{1}$.

Given the interface value of the Green's function in the L-layer, one
finds the transfer matrix from Eqs.(\ref{b9a},\ref{c9a}), and
Eqs(\ref{d9a},\ref{e9a}) give the R-layer Green's function.  Equations
analogous to (\ref{d9a},\ref{e9a}) can be written for the L-layer.
Numerical iteration of the system of the equations, allows one to
evaluate the physical properties of the system.

Motivated by the recent idea \cite{FauBelBla99} about the paramagnetic
instability at normal-metal - superconductor interfaces in the
situation where the proximity induced order parameter in the normal
metal is negative (repulsion), we consider the case when $\Delta_{L}=
- \Delta_{R}$ (questions related to the self-consistency equation are
left aside here).  The angular resolved density of states (DOS), {\it
i.e.}  $\text{Re}\, (\hat{g}^{R})_{11}$ for the almost transparent
interface $R=0.1$ is presented in Fig. 1.  Note the zero value of DOS
at small energies: DOS is extremely sensitive to reflection which
leads to the splitting of the zero energy Andreev levels.
Accordingly, the superfluid density $\rho_{s}(T)$ ($\bbox{j} = -
const\cdot\rho_{s}(T) \bbox{A}$),

\centerline{$\rho_{s}= 1 -\int_{-\infty}^{\infty} d \varepsilon \;
 \left(- {\partial f_{0}\over{\partial\varepsilon}}\right)
 \left<\text{Re}(\hat{g}^{R})_{11}\right>$}

\noindent ($\left<... \right>$ denoting the angular and space
averaging) is also very sensitive to reflection.  As seen from Fig.2,
the low temperature paramagnetic effect ({\it i.e.} $\rho_{s}<0$)
being strong for ideal interface, $R=0$, disappears when the
probability of reflection is as low as 0.04. 
These findings confirm the
theoretical possibility of the paramagnetic effect, and at the same
time indicate that the  effect is very sensitive to the inteface
reflection inevitable in the experiment.  

In conclusion, a new method which allows one to describe a system with
multiple interfaces in the framework of the quasiclassical theory is
presented.

\begin{figure}[htbp]
\epsfig{file=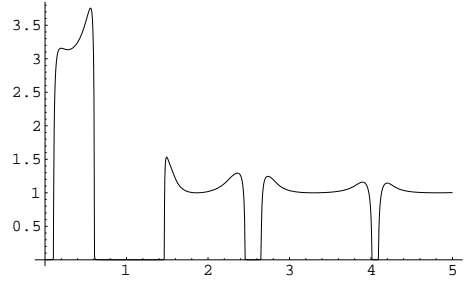,width=0.4\textwidth,angle=0}
\caption{Angle  resolved DOS vs (energy/$\Delta)$; \protect\\ 
\mbox{$\Delta_{L}=-\Delta_{R}= \Delta $}, 
$d_{L}/\cos\theta =d_{R}/\cos\theta= v/\Delta $}.
\end{figure}

\vspace{-7ex}

\begin{figure}[htbp]
\epsfig{file=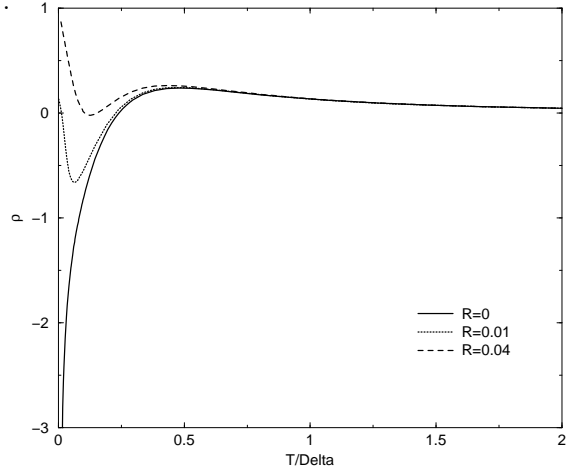,width=0.4\textwidth,angle=-90}
\caption{Space averaged superfluid density vs temperature for
the reflection  
$R=0,\;0.01,\;0.04 $}
\end{figure}

\end{document}